\documentclass[12pt]{elsarticle}
\usepackage{epstopdf}
\usepackage{amsmath}
\usepackage{cases}
\usepackage[caption=false,font=footnotesize]{subfig}
\usepackage{fixltx2e}
\usepackage{mathtools}
\usepackage{bm}
\usepackage{multirow}
\usepackage{color}
\usepackage{enumerate}
\usepackage{graphicx}
\usepackage{array}
\usepackage{threeparttable}
\usepackage{url}
\usepackage{changepage}
\usepackage[noend]{algpseudocode}
\usepackage{algorithmicx,algorithm}

\usepackage{amssymb}

\journal{Electric Power Systems Research}

\begin{document}
\begin{frontmatter}



\title{A State-Failure--Network Method to Identify Critical Components in Power Systems}

\author[zju]{Linzhi Li}
\ead{lilinzhilee@gmail.com}
\author[zju]{Hao Wu\corref{cor1}}
\ead{zjuwuhao@zju.edu.cn}
\author[zju,um]{Yonghua Song}
\ead{yhsongcn@zju.edu.cn}
\author[hn]{Yi Liu}
\ead{liuyi1@ha.sgcc.com.cn}

\cortext[cor1]{Corresponding author}
\address[zju]{College of Electrical Engineering, Zhejiang
University, Hangzhou 310027, Zhejiang, China}
\address[um]{Department of Electrical and Computer Engineering,
University of Macau, Macau, China}
\address[hn]{State Grid Henan Electric Power Company, Zhengzhou
450052, Henan, China}

\begin{abstract}
In order to mitigate cascading failure blackout risks in power systems, the critical components whose failures lead to high blackout risks should be identified.
In this paper, such critical components are identified by the state-failure network (SF-network) formed by cascading failure chain and loss data, which can be gathered from either utilities or simulations.
The failures along the chains are recombined in the SF-network, where each failure is allocated a value that can reveal the blackout risks after their occurrences.
Thus, critical failures can be identified in the SF-network where the failures raise up blackout risks, and thus the critical components can be found based on their critical failure risks.
The simulation results validate the effectiveness of the proposed method.
\end{abstract}

\end{frontmatter}

\section{Introduction}
Cascading failures in power systems often give rise to large blackouts
\cite{veloza2016analysis,guo2017critical}, which lead to severe economic losses and serious social impacts.
Generally, a cascade is triggered by one or more initial outages, and followed by many subsequent failures.
The components, whose failures are closely related to large blackouts among the subsequent failures, are the system vulnerabilities and thus more critical than other components.
An efficient way to decrease blackout risks is to upgrade the critical components, which need to be identified.
To this end, researchers have made many efforts.

Since electric grids can be taken as complex networks, some analytic methods in graph theory have been used to identify the system vulnerabilities \cite{bompard2011structural,dai2014improved,yan2014integrated,
Hines2010Do,chu2017complex}.
However, the conclusions derived from topological methods can be inaccurate because they neglect essential electrical characteristics \cite{Hines2010Do,chu2017complex}.
On the other hand, the conclusions reflect vulnerabilities only within intact systems before failure propagations, therefore, they cannot distinguish critical components under cascading failures \cite{bompard2016framework,wei2017novel}.
In fact, triggering events and propagating events have different mechanisms \cite{dobson2017cascading}.
It still calls for simulations of the cascading dynamics to analyze the critical components in the propagation phase.

Many models, such as OPA model \cite{Dobson2002An,mei2008study}, Manchester model \cite{Nedic2006Criticality}, DCSIMSEP model \cite{eppstein2012random}, stochastic model \cite{wang2012markov,rahnamay2014stochastic} and dynamic model \cite{Mousavi2012Blackouts,song2016dynamic}, have been proposed to simulate the cascading failures.
The simulation results provide the bases for analyzing the interactions between components and identifying the critical factors for cascading failure propagations \cite{hines2013dual,qi2015interaction}. 
Such interactions can be revealed by a failure-propagation--based directed graphs \cite{hines2013dual}, where nodes represent the failures, edges represent relationships between failures, and edge weights represent relationship strengths.
The interaction model in \cite{qi2015interaction} and influence graph in \cite{hines2017cascading} to identify critical components only consider event numbers and cascading sizes as the blackout severities while blackout losses are neglected.
In fact, long propagating failure sequences may not cause more severe blackouts than short ones.
The cascading failure graph in \cite{wei2017novel} considers the blackout losses but limits the cascading to propagate along the most overloaded components, ignoring the stochasticity in the propagations.
Ref. \cite{luo2017identify} shows a link analysis algorithm to analyze the importance of the components in a failure-propagation graph.

In this paper, we propose a novel method based on state-failure--network (SF-network), which is built up by cascading failure chain and loss data, to identify the critical components in the propagation phase.
The data can be gathered from either utilities or simulations.
To form the SF-network, the states that indicate failed components and failures that probably occur as subsequent events at the state are combined.
Then, values of states and failures, which can imply the expected final blackout losses, are calculated.
By identifying the critical failures in the SF-network which can raise up blackout risks, component criticality indexes can be obtained.
Therefore, components with higher indexes are more urgent to be upgraded for blackout mitigation.

Unlike the method in \cite{luo2017identify}, SF-network method calls for few assumptions or parameters.
After a plain calculation of the blackout risks of failures, it is intuitive that the failures with higher risks are more critical.
Thus, countermeasures that can curtail the occurrence probabilities of such failures can be employed to mitigate blackout risks.

The rest of the paper is organized as follows.
Section \ref{sec:SF-network} explains the detailed structure and some characteristics of the SF-network.
The identification of critical failures and components is illustrated in Section \ref{sec:identify}.
Finally, Section \ref{sec:results} studies the IEEE 39-bus system and the IEEE RTS-96 system to validate the proposed method.


\section{Structure of the State-Failure Network and Value Calculation}
\label{sec:SF-network}

\subsection{States and failures of the SF-network}
\label{sub:StatesAndFaults}
Many failure data can be recorded in real system operations or simulations, in the form of failure chains \cite{wei2017novel,hines2017cascading} with associated final losses.
Use nodes to signify the states and directed edges the failures, then an SF-network can be formed by arranging the states and failures in occurrence sequences.

Let a sequence $\lbrace{f_{(1)},\dots,f_{(k)}}\rbrace$ denote a $k$-failure chain in a $n$-component system. 
Along the chain, $k+1$ states can be set as $\lbrace{\bm{s}_0},\bm{s}_1,\dots,\bm{s}_k\rbrace$.
Each state is denoted as a 0-1 vector, whose dimension equals to $n$.
The vectors comprise zeros standing for the operating components and ones for the failed \cite{wang2012markov}.

The subscripts of the vectors, which equal to the failed component numbers, define their stages.
Thus, a state vector $\bm{s}_k$, which contains $k$ ones and $n-k$ zeros, introduces that $k$ components has failed and the state is at stage $k$.

Then the failures are combined with associated states in tuples as $(\bm{s}_0,f_{(1)})$, $(\bm{s}_1,f_{(2)})$, $\dots$, $(\bm{s}_{k-1},f_{(k)})$.
Define an ending mark which is denoted as $f_0$ (equals to 0 in this paper), then combine it with the final state in a tuple as $(\bm{s}_{k},f_0)$.
Hence, the failure chain can be rewritten as a series of state-failure tuples, and the tuples indicate where the failures are in the SF-network.
The failures join the states and the occurrence of a failure change the system from a state to the next.
Consequently, the historical propagation paths of cascading failures can be included.

\begin{figure}[tb]
	\centering
		\includegraphics[width=3.5in]{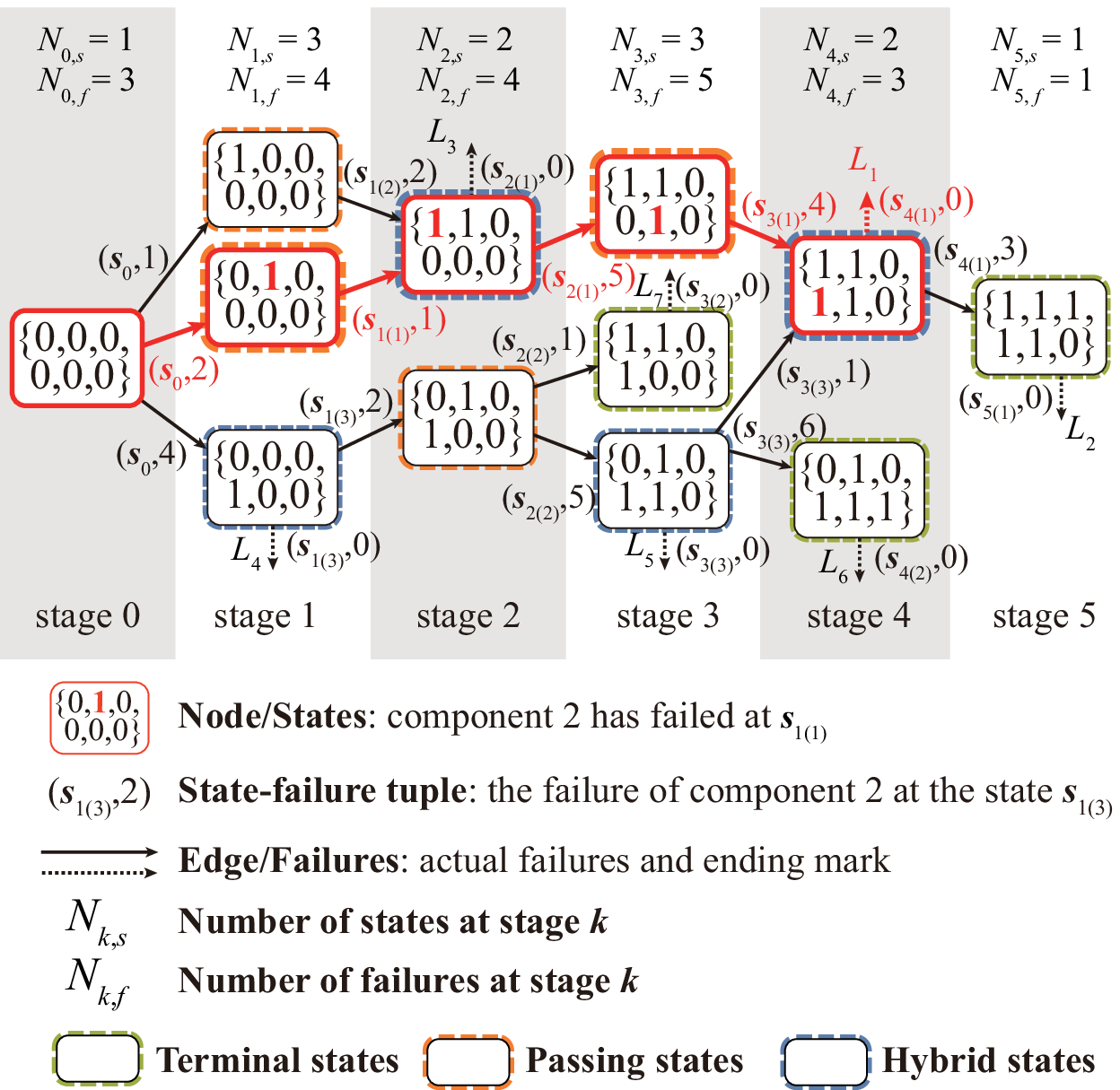}
	\caption{An example SF-network framework of a 6-component system, with data $\lbrace{2,1,5,4}\rbrace\!\to\!{L_1}$ (red color nodes and edges), $\lbrace{4,2,5,1,3}\rbrace\!\!\to\!\!{L_2}$, $\lbrace{1,2}\rbrace\!\to\!{L_3}$, $\lbrace{4}\rbrace\!\!\to\!\!{L_4}$, $\lbrace{4,2,5}\rbrace\!\!\to\!\!{L_5}$, $\lbrace{4,2,5,6}\rbrace\!\!\to\!\!{L_6}$ and $\lbrace{4,2,1}\rbrace\!\!\to\!\!{L_7}$.}
	\label{fig:stateExample}
\end{figure}

Say there is a failure chain $\lbrace{2,1,5,4}\rbrace$ gathered from a 6-component system, which can be rewritten as $(\bm{s}_0,2)$, $(\bm{s}_{1(1)},1)$, $(\bm{s}_{2(1)},5)$ and $(\bm{s}_{3(1)},4)$ and $(\bm{s}_{4(1)},0)$.
This chain as well as another 6 chains forms an SF-network in Fig. \ref{fig:stateExample}, where the symbol $\bm{s}_{k(j)}$ means the $j$th state vector at stage $k$.

The loss $L$ is joined to the end of a chain as its result, like $\lbrace{f_{(1)},\dots,f_{(k)}}\rbrace\to{L}$.
In the SF-network, losses are coupled with the state-failure tuples containing $f_0$, which are signified by the dashed directed edges in Fig. \ref{fig:stateExample}.

Here we assume that cascading failures are Markovian \cite{wang2012markov,rahnamay2014stochastic,hines2017cascading}, so the failures in SF-network only depend on the current system states.
The value calculation and more characteristics based on this assumption will be discussed in Section \ref{sub:Values}.

Then three terms for the SF-network are explained below:
\vspace*{-9pt}
\subsubsection*{1) \textbf{Probable failures} at a state}

Say a state $\bm{s}_{k(j)}$, then there can be many state-failure tuples containing $\bm{s}_{k(j)}$.
The failures in the tuples are thus the probable failures at $\bm{s}_{k(j)}$, and they are collectively originating from $\bm{s}_{k(j)}$ in the SF-network.
Particularly, an SF-network always originates from the initial state $\bm{s}_0$.
Hence, the failures at $\bm{s}_0$ are the probable N-1 events.

Note that the failures of the same component may occur at different states, they can be distinguished by state-failure tuples, such as $(\bm{s}_0,1)$, $(\bm{s}_{1(1)},1)$ and $(\bm{s}_{2(2)},1)$ showing the failures of component 1 at states $\bm{s}_0$, $\bm{s}_{1(1)}$ and $\bm{s}_{2(2)}$ respectively in Fig. \ref{fig:stateExample}.

\vspace*{-9pt}
\subsubsection*{2) \textbf{Prior states and failures} of a state}

Say $S_{\bm{s},\bm{s}_{k(j)}}$ be the set of states at stage $k-1$ which are linked to state $\bm{s}_{k(j)}$, then the states $\bm{s}^{k(j)}_{(i)}\in S_{\bm{s},\bm{s}_{k(j)}}$ are taken as the prior states of $\bm{s}_{k(j)}$.
Each $\bm{s}^{k(j)}_{(i)}$ is linked to $\bm{s}_{k(j)}$ by associated failures $f^{k(j)}_{(i)}\in S_{f,\bm{s}_{k(j)}}$, where $f^{k(j)}_{(i)}$ are taken as the prior failures of $\bm{s}_{k(j)}$ and $S_{f,\bm{s}_{k(j)}}$ is the set of prior failures.

In Fig. \ref{fig:stateExample}, for example, $\bm{s}_{1(1)}$ and $\bm{s}_{1(2)}$ are the prior states of $\bm{s}_{2(1)}$, and failures of components 1 and 2 are the prior failures of $\bm{s}_{2(1)}$ .

\vspace*{-9pt}
\subsubsection*{3) \textbf{Types of the states}}

There are three types of states: terminal state with only $f_0$ where a cascade definitely ends, passing state without $f_0$ where a cascade definitely propagates to the next stage, and hybrid state with actual failures and $f_0$ where a cascade can end or pass.
The terminal, passing and hybrid states are surrounded by green, orange and blue dashed circles respectively in Fig. \ref{fig:stateExample}.


\subsection{Values of states and failures in the SF-network}
\label{sub:Values}
In addition to the forming process above, loss data are used to calculate the values of states and failures in the SF-network.

Here we define the calculation rules, where the state values and failure values are worked out from the last to the first stage.
For simplicity, the state values and failure values are abbreviated by S-values and F-values hereafter.

Assume that the F-values of the failures $f_m$ at a state $\bm{s}_{k(j)}$ are known, where the subscript $m$ represents the $m$th component in the system.
Let $S_{\bm{s}_{k(j)}}$ be the set of the probable failures at $\bm{s}_{k(j)}$.
Since the occurrences of the failures are Markovian, the S-value of the state is obtained by 
\begin{equation}
\begin{split}
S(\bm{s}_{k(j)})=&\!\!\!\!\sum_{f_m\in{S_{\bm{s}_{k(j)}}}}\!\!\!\!\frac{N_{(\bm{s}_{k(j)},f_m)}}{N_{\bm{s}_{k(j)}}}
F(\bm{s}_{k(j)},f_m)\\
=&\!\!\!\!\sum_{f_m\in{S_{\bm{s}_{k(j)}}}}\!\!\!\!p(f_m|\bm{s}_{k(j)})F(\bm{s}_{k(j)},f_m),
\end{split}
\label{equ:stateValueCal}
\end{equation}
where $N_{\bm{s}_{k(j)}}$ and $N_{(\bm{s}_{k(j)},f_m)}$ are the occurrence counts of the state and failure according to the data, and
\begin{equation}
N_{\bm{s}_{k(j)}}=\!\!\!\!\sum_{f_m\in{S_{\bm{s}_{k(j)}}}}\!\!\!\!N_{(\bm{s}_{k(j)},f_m)}.\label{equ:countEqual}
\end{equation}
Thus, $p(f_m|\bm{s}_{k(j)})$ is the conditional probability of $f_m$ at $\bm{s}_{k(j)}$.

The F-values in (\ref{equ:stateValueCal}) are derived from two aspects.
For the state-failure tuples containing $f_0$ where the losses are firstly introduced into the SF-network, their F-values equal to the losses
\begin{equation}
F(\bm{s}_{k(j)},f_0)=L,\quad f_0\in{S_{\bm{s}_{k(j)}}}.\label{equ:Loss2Fvalue}
\end{equation}
For the other tuples whose failures head for states at the next stage, their F-values equal to the S-values of the states
\begin{equation}
F(\bm{s}_{k(j)},f_m)=S(\bm{s}_{k+1(j')}),\quad f_m\in{S_{\bm{s}_{k(j)}}},\label{equ:Svalue2Fvalue}
\end{equation}
where $\bm{s}_{k(j)}$ and $f_m$ are prior states and failures of $\bm{s}_{k+1(j')}$.

\begin{figure}[tb]
	\centering
		\includegraphics[width=3.3in]{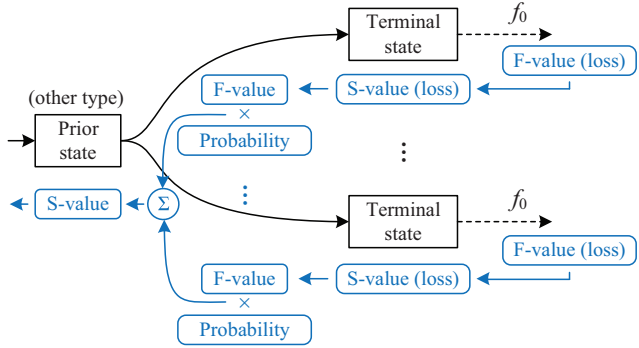}
	\caption{The calculation relationships of the S-values and F-values. The black part is the SF-network structure, and the blue part is the calculation flow charts. Obviously, the value calculation can be regarded as a upstream transmission process, where losses are transmitted from the terminal states to prior states.}
\label{fig:ValueCalculation}
\end{figure}

Therefore, though ending marks $f_0$ are not actual failures that are recorded in failure chains, they introduce the losses into the SF-network and play the same role like actual failures in (\ref{equ:stateValueCal}), (\ref{equ:Loss2Fvalue}) and (\ref{equ:Svalue2Fvalue}), and thus can be seen as fake failures.
Fig. \ref{fig:ValueCalculation} briefly depicts the relationships between the values of states and failures, which shows the upstream transmission characteristic of the value calculation.

Before the calculations, all values of the states and failures in an SF-network are unknown except for those of the fake failures.
So, the calculations start at the terminal state at the last stage based on (\ref{equ:Loss2Fvalue}), and the S-value of the terminal state equals to the only F-value.
Then all the prior failures of the terminal states get their F-values based on (\ref{equ:Svalue2Fvalue}).
Continue calculating at prior states with multiple failures based on (\ref{equ:stateValueCal}) until to the first stage, then all values of the SF-network are worked out.

After the calculations, each state and failure gets a value.
As shown in Fig. \ref{fig:ValueCalculation}, the S-values of terminal states are the loss averages of the failure chains that end here.
Therefore, the F-values of prior failures indicate the expected final losses after their occurrences, although it is unknown whether intermediate loss appears after this failure occurs.
Thus, the products of these values and the probabilities are equivalent to the blackout risks \cite{vaiman2012risk}. 
In other words, the F-values times the conditional probabilities indicate the final blackout risks of the failure at a state.
Then, the S-value quantifies the blackout risk of the system at the state according to (\ref{equ:stateValueCal}).

\subsection{Conservation of values of failure cutsets}
\begin{figure}[tb]
	\centering
		\includegraphics[width=2.4in]{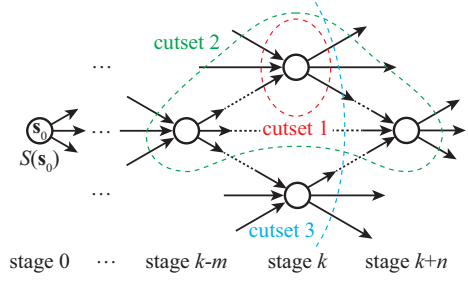}
	\caption{The dashed curves are failure cutsets in the SF-network. Nodes are states and directed edges are failures.}
\label{fig:cuts}
\end{figure}

Denote failures that cut the SF-network into separate networks as a failure cutset, the failures that head into the cutset as inward-heading failures, and the failures that head outside as outward-heading failures, as the cutsets in Fig. \ref{fig:cuts}.
Then the value calculation in SF-network manifests the conservation of values of failure cutsets.

Firstly, from (\ref{equ:stateValueCal}) we can get
\begin{equation}
N_{\bm{s}_{k(j)}}S(\bm{s}_{k(j)})=\!\!\!\!
\sum_{f_m\in{S_{\bm{s}_{k(j)}}}}\!\!\!\!N_{(\bm{s}_{k(j)},f_m)}F(\bm{s}_{k(j)},f_m),
\label{equ:outValueSum}
\end{equation}
which means that S-value of the state $\bm{s}_{k(j)}$ times its occurrence count equals to the sum of F-values of probable failures times occurrence counts.
Let ``value sum'' be short for the sum of values times occurrence counts.
Then, similar to (\ref{equ:countEqual}), it is easy to know that total occurrence count of the prior failures of $\bm{s}_{k(j)}$ equals to $N_{\bm{s}_{k(j)}}$.
Thus, given that each F-value of prior failures equals to the S-value based on (\ref{equ:Svalue2Fvalue}), the equation between value sum of state and its prior failures can also be derived from
\begin{equation}
N_{\bm{s}_{k(j)}}S(\bm{s}_{k(j)})=
\!\!\!\!\!\!\!\!\sum_{f^{k(j)}_{(i)}\in S_{f,\bm{s}_{k(j)}}}
\!\!\!\!\!\!\!N_{(\bm{s}^{k(j)}_{(i)},f^{k(j)}_{(i)})}
F(\bm{s}^{k(j)}_{(i)},f^{k(j)}_{(i)}).\label{equ:inValueSum}
\end{equation}

Assume a cutset surrounds a single state like cutset 1 in Fig. \ref{fig:cuts}, then the outward-heading failures are the probable failures and the inward-heading ones are the prior failures.
Hence, value sum of inward-heading failures equals to that of outward-heading failures base on (\ref{equ:outValueSum}) and (\ref{equ:inValueSum}).

Generally, a cutset can surround multiple states.
If it excludes the initial state $\bm{s}_0$, such as cutset 2 in Fig. \ref{fig:cuts}, which is actually a union set of the cutsets surrounding the inner single states.
The value sum equation also holds for the failures in these failure cutsets, which is
\begin{equation}
\begin{split}
E_\mathrm{cutset}=&\sum N_{(\cdot,f_\mathrm{in})}F(\cdot,f_\mathrm{in})\\
&-\Big(\sum N_{(\cdot,f_\mathrm{out})}F(\cdot,f_\mathrm{out})+\sum N_{(\cdot,f_0)}F(\cdot,f_0)\Big)\\
=&E_\mathrm{in}-E_\mathrm{out}=0,
\end{split}\label{equ:valueConservation}
\end{equation}
where $E_\mathrm{cutset}$ is the value sum of the failures in a cutset, and $f_\mathrm{in/out}$ denotes the inward/outward-heading failures.
The fake failures may not be in the cutset, their value sum is counted in the $E_\mathrm{out}$ part.
Thus, (\ref{equ:valueConservation}) reveals the conservation of values in cutsets that surround multiple states without the initial state.
Note that the states surrounded by a cutset are not necessarily connected.


Now, to further expand (\ref{equ:valueConservation}), the S-value item of $\bm{s}_0$ needs to be added to (\ref{equ:valueConservation}) when a failure cutset surrounds initial state $\bm{s}_0$,
Thus, we obtain
\begin{equation}
E_\mathrm{cutset}=NS(\bm{s}_0)+E_\mathrm{in}-E_\mathrm{out}=0,
\label{equ:valueConservation2}
\end{equation}
where $N$ is the total number of the failure chains.
Then for the cutset without inward-heading failures, such as cutset 3 in Fig. \ref{fig:cuts}, $E_\mathrm{in}=0$ in (\ref{equ:valueConservation2}).
Thus, (\ref{equ:valueConservation2}) shows the value conservation in failure cutsets in an SF-network, where the value sums of any failure cutset equal to 0.

As a special case, when the cutset is so large that all failures are surrounded by the cutset, (\ref{equ:valueConservation2}) becomes
\begin{equation}
E_\mathrm{cutset}=NS(\bm{s}_0)-\sum N_{(\cdot,f_0)}F(\cdot,f_0)=0, \label{equ:valueConservation3}
\end{equation}
which says that the S-value sum of $\bm{s}_0$ equals to the sum of all losses, and indicates that the value upstream transmission calculations through the SF-network are lossless.

\section{Identify Critical Components with the SF-Network}
\label{sec:identify}
\subsection{Identify the critical failures in the SF-network}
\label{sub:criticalFailures}
\begin{figure}[tb]
	\centering
		\includegraphics[width=2.4in]{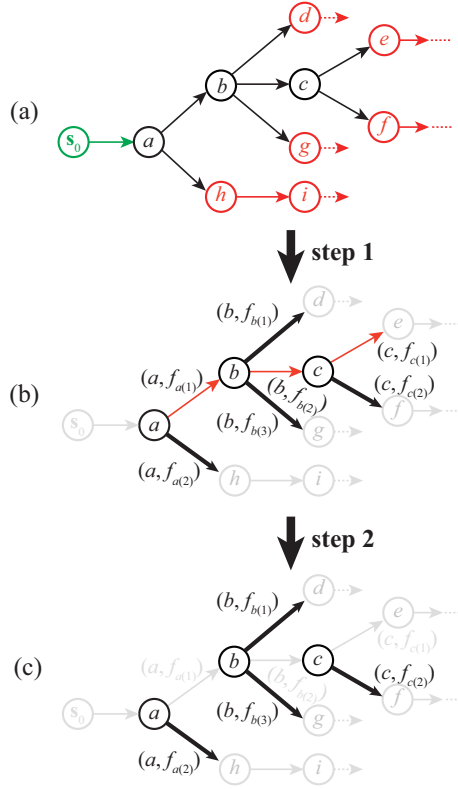}
	\caption{Illustration of identifying the critical failures in the SF-network.}
	\label{fig:trimFailure}
\end{figure}

According to Section \ref{sub:Values}, the S-value quantifies the blackout risk of the system at a state, and the F-value indicates the blackout risk after a failure occurs.
Thus, in a complete SF-network formed by sufficient data, the failures whose F-values are higher than the S-values of associated states are worth more attention, because their occurrences can increase system blackout risks.
Therefore, we call such failures critical failures in the SF-network, and identify them by the process as shown in Fig. \ref{fig:trimFailure}.

Firstly, exclude the non-critical failures based on the structural features of the SF-network in step 1.
There are single failures at some states, as the red part in Fig. \ref{fig:trimFailure}(a), which can be seen as the results of prior failures.
Thus, the forked structures in the SF-network are remained.

Because the occurrence mechanisms differ between initiating events, which are mainly caused by outer factors including lightning stroke and tree contacts, and subsequent events, which are the system responses caused by overcurrent relays or operator orders \cite{dobson2017cascading,hines2017cascading}.
Here in this paper, we focus on the critical subsequent events in the propagation phase, and the initiating events, which are colored in green in Fig. \ref{fig:trimFailure}(a), are not considered. 

Then, among the multiple failures, the failures whose F-values are no more than the S-values of the states are excluded in step 2.
Assume the bold part in Fig. \ref{fig:trimFailure}(b) are the failures whose F-values are higher than the S-values, then they are the identified critical failures, as the failures in tuples $(a,f_{a(2)})$, $(b,f_{b(1)})$, $(b,f_{b(3)})$ and $(c,f_{c(2)})$ in Fig. \ref{fig:trimFailure}(c).

\subsection{Critical failure risk and identify the critical components}
\label{sub:identify}
The identified failures are critical at their own states, to evaluate and compare the risks of failures at different states and stages, the nonconditional probabilities need to be worked out.

Let $N_k$ be the times that stage $k$ is reached.
Then the nonconditional probability of a failure can be derived from

\begin{equation}
\begin{split}
p(\bm{s}_{k(j)},f_m)&=\frac{N_k}{N}
\frac{N_{\bm{s}_{k(j)}}}{N_k}
p(f_m|\bm{s}_{k(j)})\\
&=\frac{N_{(\bm{s}_{k(j)},f_m)}}{N}.
\end{split}\label{equ:GlobalProb}
\end{equation}
Thus, given that F-values are the expected losses of failures as illustrated in Section \ref{sub:Values}, the risk of a failure at any state can be expressed by
\begin{equation}
\begin{split}
R(\bm{s}_{k(j)},f_m)&=p(\bm{s}_{k(j)},f_m)F(\bm{s}_{k(j)},f_m)\\
&=\frac{N_{(\bm{s}_{k(j)},f_m)}}{N}F(\bm{s}_{k(j)},f_m).
\end{split}
\end{equation}
Therefore, after the items in (\ref{equ:valueConservation2}) and (\ref{equ:valueConservation3}) are divided by $N$, the value conservation turns into the risk conservation of failure cutsets.

Sum up the critical failure risks of a component, the component criticality index (CCI) can be obtained from
\begin{equation}
R_m=\sum\limits_{k=k_0}^\infty\sum\limits_{j=1}^{N_k}
\frac{N_{(\bm{s}^*_{k(j)},f^*_m)}}{N}F(\bm{s}^*_{k(j)},f^*_m),
\label{equ:compRiskIdentified}
\end{equation}
where $k_0$ is the stage right after initiating events, and the asterisk means that the variables are the identified critical failures in the SF-network.
Note that $R_m$ is not strictly equivalent to component failure risks due to that some non-critical failures are excluded in (\ref{equ:compRiskIdentified}).
Therefore, CCI is an assessing index for component criticality, and the components with higher CCI are more critical.

With a complete SF-network where critical failures can be accurately identified, the component criticality results are reliable.
However, it is impractical to gather all failure chains, and thus impossible to achieve thorough completeness of the SF-network.
The failures that are too rare to affect the risk assessments can be neglected, that is, gathering a majority of cascading failures can lead to reliable criticality analyses.
As failure chain data are read in and the SF-network is evolving, if CCIs become stable enough (the changes of CCIs converge), the failure chain data will be sufficient to identify critical components by the SF-network.

\section{Simulation Results}
\label{sec:results}
\subsection{Cascading failure simulator}
Among the cascading failure models, the static models are the most popular due to their computational efficiency and reliable analytical results.
Although the cascading failure processes are complicated, the transmission outages are the dominant forces in the propagation phase.
Thus, static models, such as OPA model, are capable to simulate the behaviours of the transmission network in the cascading failure through the load redistribution mechanism.
Here we design a DC power flow based simulator to produce necessary cascading failure chains and validate the method, and the components in this paper are branches.
The procedures are as below:
\begin{enumerate}[Step 1:]
\item Input the power system initial operation point.
\item Set initiating contingency to trigger a cascading failure.
\item Detect islands. Rebalance the generation and loads by generator rampings with participation factors in each islands. Trip generators or shed loads when generator rampings cannot cover the power mismatch \cite{eppstein2012random,rezaei2016cascading}.
    Go to step \ref{stp:powerflowCal}.\label{stp:rebalance}
\item Calculate DC power flow and check whether any load flow of branches exceeds its capacity $F_{i,\mathrm{c}}$. If yes, go to step \ref{stp:tripLine}, if not, go to step \ref{stp:lossCount}. \label{stp:powerflowCal}
\item Select a branch to trip based on the probability model \cite{jia2016risk,luo2017identify} below and append this failure to the failure chain
\begin{equation}
p_{i,\mathrm{trip}}=\begin{cases}
   1\!\!\!\!\!&,F_i\ge F_{i,max}\\
   \frac{F_i-F_{i,\mathrm{c}}}{F_{i,\mathrm{max}}-F_{i,\mathrm{c}}}\!\!\!\!\!&,F_{i,c}<F_i<F_{i,\mathrm{max}} \\
   0\!\!\!\!\!&,F_i\le F_{i,\mathrm{c}}
   \end{cases}\!\!,\label{equ:overloadTrip}
\end{equation}
where $F_{i,\mathrm{max}}$ is the capacity limit of $i$th branch.
When multiple branches are probable to outage derived from (\ref{equ:overloadTrip}), randomly select a branch among them to trip \cite{wei2017novel,rahnamay2014stochastic}. If line tripping happens, go to step \ref{stp:rebalance}.
Otherwise, go to step \ref{stp:lossCount}. \label{stp:tripLine}
\item The cascade ends. Record the cascading failure chain, and count the blackout loss as the result of the failure chain.\label{stp:lossCount}
\end{enumerate}

\subsection{IEEE 39-bus system}
The system data can be accessed from \cite{MatpowerSystemData}.
It has 46 branches and 6254.23MW load.
$F_{i,\mathrm{max}} = 1.4F_{i,c}$.
Load losses are taken as the losses in the SF-network and the values are normalized and divided by the total load amount.
Random N-2 contingencies are chosen to trigger cascading failures.

\begin{figure}[tb]
	\centering
		\includegraphics[width=3.5in]{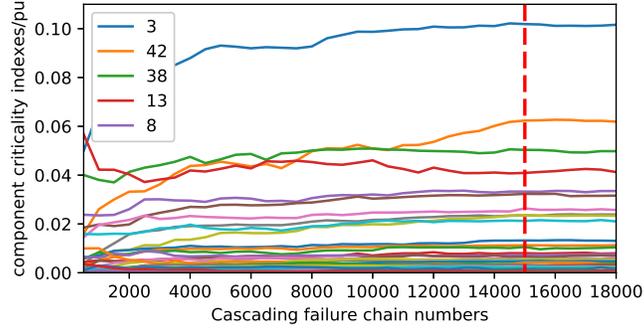}
	\caption{The convergence of the CCIs of the IEEE 39-bus system.}
	\label{fig:39converge}
\end{figure}

\begin{figure}[tb]
	\centering
		\includegraphics[width=3.5in]{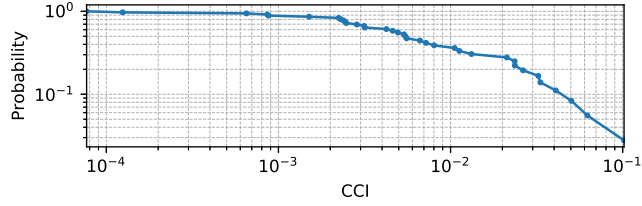}
	\caption{The CCD of the CCIs of the IEEE 39-bus system.}
	\label{fig:39ccd}
\end{figure}

As the failure chains form the SF-network, the CCIs are recorded as shown in Fig. \ref{fig:39converge}.
Apparently, the CCIs level after about 15000 chains (right side of the red dashed line) and reach convergence.
We calculate the CCIs in the SF-network formed by the 15000 failure chains.
The complementary cumulative distribution (CCD) of the CCIs are given in Fig. \ref{fig:39ccd}.
As seen in the loglog plot, the CCIs show a power-law distribution, which indicates that only a small part of the branches are much more critical than others.
Qualitatively similar results of the component criticality distributions have also been obtained in \cite{wei2017novel,qi2015interaction,hines2017cascading}.


The identification results are validated by branch capacity expansion \cite{hines2017cascading,luo2017identify,wenli2016cascading}, which can be accomplished by reconducting or by adding parallel transmission corridors in real-life systems \cite{hines2017cascading}.
The expansion can be easily modeled as
\begin{equation}
\begin{cases}
F'_{i,\mathrm{c}}=F_{i,\mathrm{c}} +\Delta{F}\\
F'_{i,\mathrm{max}}=F_{i,\mathrm{max}} +\Delta{F}
\end{cases},
\end{equation}
where $\Delta{F}$ is the expansion capacity.
Branches with expanded capacities tend to fail less frequently, which alleviates the cascading propagations.
Thus, upgrading the branches with higher CCIs should result in more effective blackout risk mitigation.
Here the capacities of the branches in each group are expanded by 500MW.

\begin{figure}[tb]
	\centering
		\includegraphics[width=3.5in]{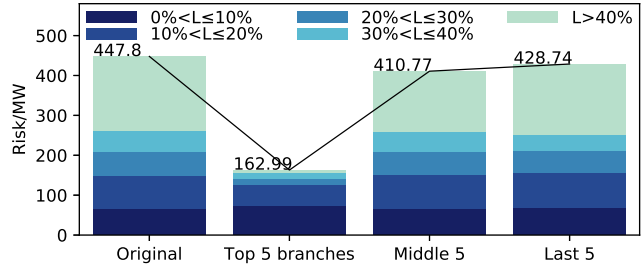}
	\caption{The blackout risks with different branches upgraded of the IEEE 39-bus system.}
	\label{fig:39selfRisk}
\end{figure}

After listing the branches in the descending order of their CCIs, we select 3 groups of branches and each group comprise 5 branches (10\% of the total 46 branches).
As shown in Table \ref{tb:39CCIresult}, the top 5 branches rank from 1st to 5th, the middle from 18th to 24th and the last from 33rd to 37th.
There are 9 out of 46 branches which do not fail in the cascading propagation phase, thus only 37 branches are included in the full list.
The systems with different upgraded branches are simulated, and the blackout risk results are obtained and shown in Fig. \ref{fig:39selfRisk}.

\begin{table}
\caption{CCIs of the IEEE 39-bus system ($\times{10^{-1}}$)}
\centering
\begin{tabular}
{m{0.8in}<{\centering}m{0.5in}<{\centering}|
m{0.8in}<{\centering}m{0.5in}<{\centering}|
m{0.8in}<{\centering}m{0.8in}<{\centering}}
\hline
\hline
\multicolumn{2}{c|}{\footnotesize Top 5 branches}&
\multicolumn{2}{c|}{\footnotesize Middle 5 branches}&
\multicolumn{2}{c}{\footnotesize Last 5 branches}\\
\hline
    \footnotesize Branch   &  \footnotesize CCI  &   \footnotesize Branch    &  \footnotesize CCI  &    \footnotesize Branch     &    \footnotesize CCI\\
\hline
\footnotesize  3(2-3)  & \footnotesize 1.019  & \footnotesize 22(12-13)  & \footnotesize 0.056  &  \footnotesize 11(5-8)   & \footnotesize 0.876$\times{10^{-2}}$  \\
\footnotesize  42(26-27) & \footnotesize 0.624  & \footnotesize 25(15-16)   & \footnotesize 0.055  &  \footnotesize 15(7-8)   & \footnotesize 0.861$\times{10^{-2}}$  \\
\footnotesize  38(23-24)  & \footnotesize 0.503  & \footnotesize 30(17-18)   & \footnotesize 0.054  &  \footnotesize 36(22-23) & \footnotesize 0.651$\times{10^{-2}}$  \\
\footnotesize  13(6-11)  & \footnotesize 0.409  & \footnotesize 21(12-11)  & \footnotesize 0.050  &  \footnotesize 45(28-29)   & \footnotesize 0.124$\times{10^{-2}}$  \\
\footnotesize  8(4-5)  & \footnotesize 0.333  & \footnotesize 24(14-15)   & \footnotesize 0.046  &  \footnotesize 44(26-29)  & \footnotesize 0.077$\times{10^{-2}}$  \\
\hline
\hline
\end{tabular}\label{tb:39CCIresult}
\end{table}

It is obvious that branch upgrade can decrease the blackout risks, which are decreased from 447.8MW to 162.99MW, 410.77MW and 428.74MW by upgrading the top, middle and last ranking branches respectively.
And upgrading the top ranking, i.e., the most critical, branches shows the largest risk decrease among the three groups, which validates the effectiveness of the proposed method.
In addition, upgrading the top ranking 5 branches can substantially decrease the risks of large blackouts (load losses higher than 20\%), while the small blackouts are scarcely affected.

\begin{figure}[tb]
	\centering
		\includegraphics[width=3.5in]{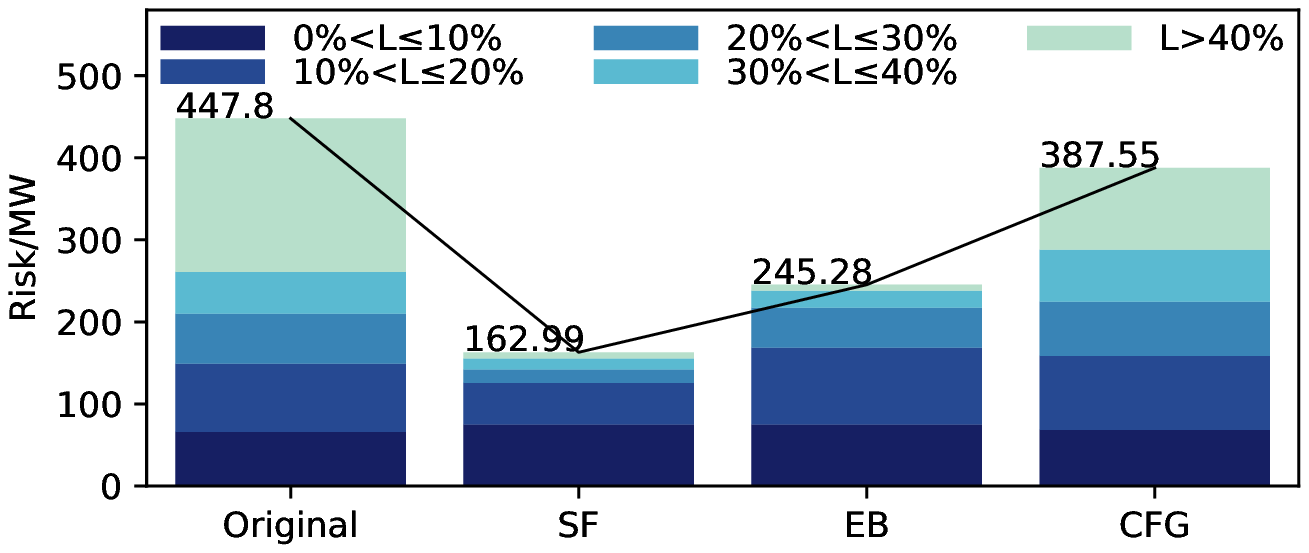}
	\caption{The blackout risks after upgrading the critical branches identified by SF-network (SF), extended betweenness (EB) and cascading failure graph (CFG) of the IEEE 39-bus system.}
	\label{fig:39crossRisk}
\end{figure}

Besides, we compare the results obtained from the proposed SF-network method (SF) with the identification results from extended betweenness (EB) method \cite{bompard2012extended} and cascading failure graph (CFG) method \cite{wei2017novel}.
Table \ref{tb:39cross} gives the top ranking 5 branches of the results.
Only the most critical branch (branch 3(2-3)) identified by SF is also contained in the EB identification result, while other branches are different. 
Fig. \ref{fig:39crossRisk} shows that the blackout risks of systems decrease from 447.8MW to 162.99MW by SF identification results, to 245.28MW by EB identification results and to 387.55MW by CFG identification results.
It is observed that upgrading the critical branches identified by SF-network can decrease blackout risks more significantly than those by EB and CFG methods.
\begin{table}
\caption{Critical branches of the IEEE 39-bus system identified by different methods}
\centering
\begin{tabular}
{c|c|c|c}
\hline
\hline
\footnotesize Rank & \footnotesize SF & \footnotesize EB & \footnotesize CFG \\
\hline
\footnotesize 1&\footnotesize3(2-3)    &  \footnotesize26(16-17)   &  \footnotesize27(17-27)    \\
\footnotesize 2&\footnotesize42(26-27) & \footnotesize 25(15-16)   & \footnotesize 20(10-32)  \\
\footnotesize 3&\footnotesize38(23-24) & \footnotesize3(2-3)      &  \footnotesize37(22-35)\\
\footnotesize 4&\footnotesize13(6-11)  & \footnotesize 27(16-19)   &  \footnotesize46(29-38)\\
\footnotesize 5&\footnotesize8(4-5)    & \footnotesize 31(17-27)   &  \footnotesize5(2-30)\\
\hline
\hline
\end{tabular}\label{tb:39cross}
\end{table}

\subsection{IEEE RTS-96 system}
The SF-network method is also validated by the case study on the IEEE RTS-96 system, whose data can be found in \cite{grigg1999ieee}.
It has 120 branches and 8550MW load.
The loads are increased by 60\% to increase the operation stress.
The initial operation point is determined by optimal power flow calculation.
And similar to the IEEE 39-bus system, $F_{i,\mathrm{max}} = 1.4F_{i,c}$.
Load losses are taken as the losses in the SF-network and the values are normalized and divided by the total load amount.
random N-2 contingencies are chosen to trigger cascading failures.

\begin{figure}[tb]
	\centering
		\includegraphics[width=3.5in]{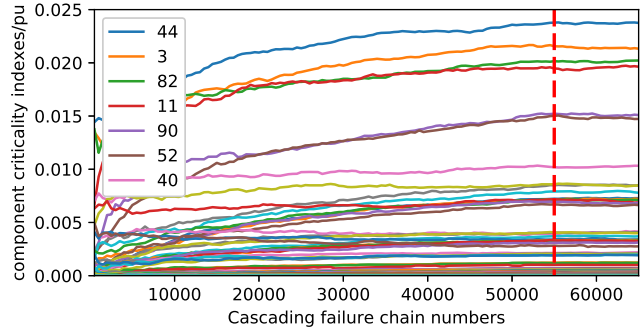}
	\caption{ The convergence of the CCIs of the IEEE RTS-96 system.}
	\label{fig:96converge}
\end{figure}

\begin{figure}[tb]
	\centering
		\includegraphics[width=3.5in]{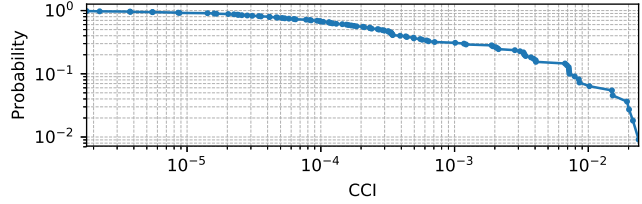}
	\caption{The CCD of the CCIs of the IEEE RTS-96 system.}
	\label{fig:96ccd}
\end{figure}

As the failure chains form the SF-network, the CCIs are recorded as shown in Fig. \ref{fig:96converge}, where CCIs level after about 55000 chains.
Thus, we use the CCIs derived from the SF-network formed by the 55000 chains, and the CCD of the CCIs shown in Fig. \ref{fig:96ccd} also show a power-law distribution.

\begin{figure}[tb]
	\centering
		\includegraphics[width=3.5in]{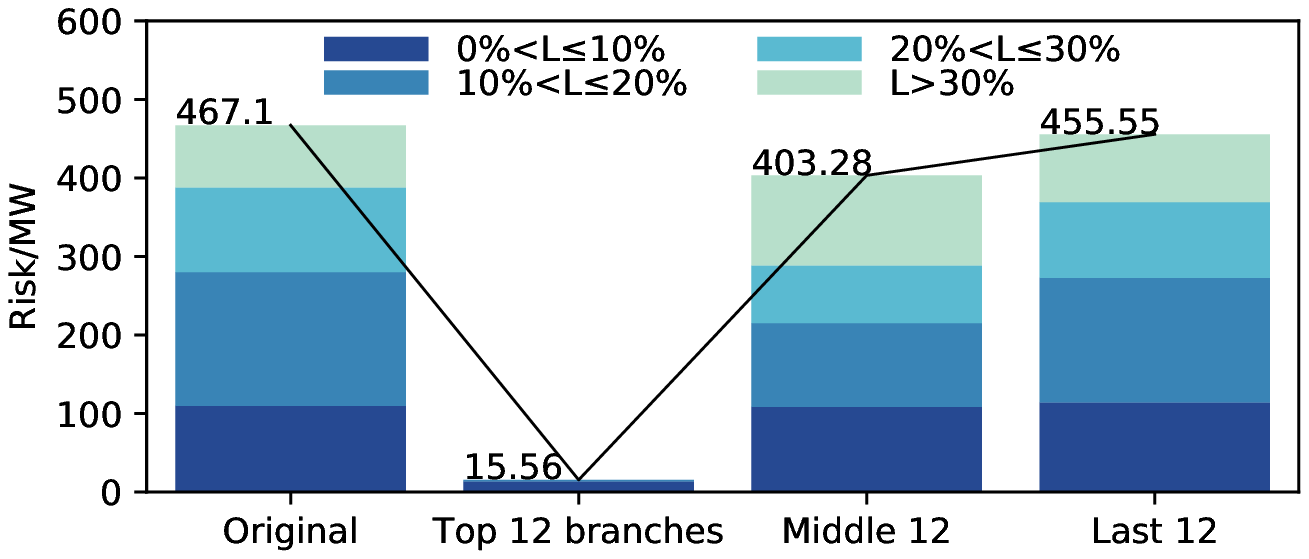}
	\caption{The blackout risks with different branches upgraded of the IEEE RTS-96 system.}
	\label{fig:96selfRisk}
\end{figure}

Similarly, select three groups of branches according to their CCI values to validate the identification results, as given in Table \ref{tb:96CCIresult}.
Here each group comprises 12 branches (10\% of the total 120 branches).
The top 12 branches rank from 1st to 12th, the middle from 27th to 38th and the last from 97th to 108th.
There are 12 out of 120 branches which do not fail in the cascading propagation phase, thus only 108 branches are included in the full list.
To upgrade the branches, their capacities are also expanded by 500MW.
Then, the systems with different upgraded branches are simulated, and the blackout risk results are shown in Fig. \ref{fig:96selfRisk}.


\begin{table}
\caption{CCIs of the IEEE RTS-96 system ($\times{10^{-1}}$)}
\centering
\begin{tabular}
{m{0.8in}<{\centering}m{0.5in}<{\centering}|
m{0.8in}<{\centering}m{0.5in}<{\centering}|
m{0.8in}<{\centering}m{0.7in}<{\centering}}
\hline
\hline
\multicolumn{2}{c|}{\footnotesize Top 12 branches}&
\multicolumn{2}{c|}{\footnotesize Middle 12 branches}&
\multicolumn{2}{c}{\footnotesize Last 12 branches}\\
\hline
    \footnotesize Branch   &  \footnotesize CCI  &   \footnotesize Branch    &  \footnotesize CCI  &    \footnotesize Branch     &    \footnotesize CCI\\
\hline
 \scriptsize 44(201-205) & \scriptsize 0.240 & \scriptsize 85(303-309) & \scriptsize 0.022 & \scriptsize 9(105-110) & \scriptsize 0.186$\times{10^{-3}}$\\
 \scriptsize 3(101-105) & \scriptsize 0.219 & \scriptsize 51(206-210) & \scriptsize 0.021 & \scriptsize 77(220-223) & \scriptsize 0.182$\times{10^{-3}}$\\
 \scriptsize 82(301-305) & \scriptsize 0.204 & \scriptsize 6(103-109) & \scriptsize 0.021 & \scriptsize 28(115-121) & \scriptsize 0.179$\times{10^{-3}}$\\
 \scriptsize 11(107-108) & \scriptsize 0.197 & \scriptsize 10(106-110) & \scriptsize 0.020 & \scriptsize 88(305-310) & \scriptsize 0.114$\times{10^{-3}}$\\
 \scriptsize 90(307-308) & \scriptsize 0.153 & \scriptsize 89(306-310) & \scriptsize 0.019 & \scriptsize 66(215-221) & \scriptsize 0.100$\times{10^{-3}}$\\
 \scriptsize 52(207-208) & \scriptsize 0.150 & \scriptsize 64(214-216) & \scriptsize 0.013 & \scriptsize 104(315-321) & \scriptsize 0.094$\times{10^{-3}}$\\
 \scriptsize 40(121-122) & \scriptsize 0.103 & \scriptsize 102(314-316) & \scriptsize 0.012 & \scriptsize 111(318-321) & \scriptsize 0.053$\times{10^{-3}}$\\
 \scriptsize 46(202-206) & \scriptsize 0.086 & \scriptsize 25(114-116) & \scriptsize 0.010 & \scriptsize 112(318-321) & \scriptsize 0.053$\times{10^{-3}}$\\
 \scriptsize 117(321-322) & \scriptsize 0.085 & \scriptsize 13(108-109) & \scriptsize 0.007 & \scriptsize 78(220-223) & \scriptsize 0.041$\times{10^{-3}}$\\
 \scriptsize 5(102-106) & \scriptsize 0.080 & \scriptsize 91(308-309) & \scriptsize 0.007 & \scriptsize 20(111-114) & \scriptsize 0.024$\times{10^{-3}}$\\
 \scriptsize 83(302-304) & \scriptsize 0.074 & \scriptsize 53(208-209) & \scriptsize 0.007 & \scriptsize 113(319-320) & \scriptsize 0.021$\times{10^{-3}}$\\
 \scriptsize 79(221-222) & \scriptsize 0.074 & \scriptsize 18(110-112) & \scriptsize 0.006 & \scriptsize 60(211-214) & \scriptsize 0.010$\times{10^{-3}}$\\
\hline
\hline
\end{tabular}\label{tb:96CCIresult}
\end{table}

As shown in Fig. \ref{fig:96selfRisk}, blackout risks decrease from 467.10MW to 15.56MW, 403.28MW and 455.55MW by upgrading top ranking, middle ranking and last ranking 12 branches respectively.
Among them, upgrading the top ranking branches eliminates the majority of cascading failures and only small blackouts may occur.
While upgrading the middle ranking and last ranking branches can achieve very limited blackout risk mitigation effectiveness.

\begin{figure}[tb]
	\centering
		\includegraphics[width=3.5in]{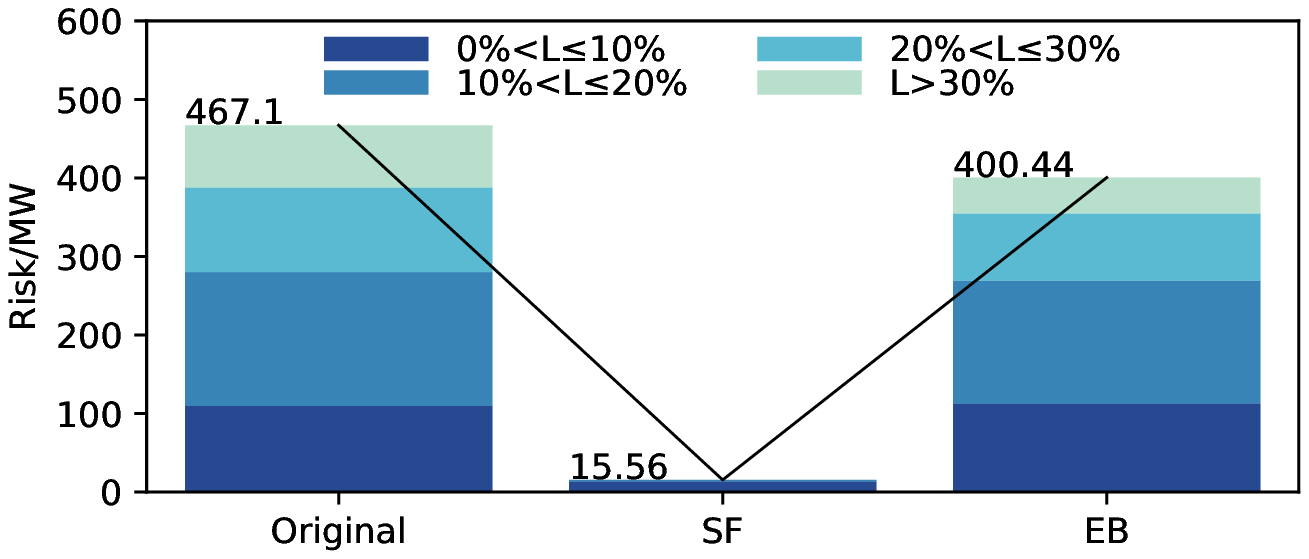}
	\caption{The blackout risks after upgrading the critical branches identified by SF-network (SF) and extended betweenness (EB) of the IEEE RTS-96 system.}
	\label{fig:96crossRisk}
\end{figure}

Then, the identification results by the proposed method are also compared with those by EB method.
The critical branches identified by the both methods are given in Table \ref{tb:96cross}, and the blackout risk contrasts are given in Fig. \ref{fig:96crossRisk}.
It can be seen that the proposed method shows better effectiveness in critical components identifications than EB method.

\begin{table}
\caption{Critical branches of the IEEE RTS-96 system identified by different methods}
\centering
\begin{tabular}
{c|c|c}
\hline
\hline
\footnotesize Rank & \footnotesize SF & \footnotesize EB \\
\hline
\footnotesize 1&\footnotesize44(201-205)    &  \footnotesize 119(318-223)  \\
\footnotesize 2&\footnotesize3(101-105) & \footnotesize   118(325-121) \\
\footnotesize 3&\footnotesize82(301-305) & \footnotesize   120(323-325) \\
\footnotesize 4&\footnotesize11(107-108)  & \footnotesize  70(216-219) \\
\footnotesize 5&\footnotesize90(307-308)    & \footnotesize 31(116-119)\\
\footnotesize 6&\footnotesize52(207-208) &  \footnotesize  65(215-216)\\
\footnotesize 7&\footnotesize40(121-122) &  \footnotesize  26(115-116)\\
\footnotesize 8&\footnotesize46(202-206) &  \footnotesize  108(316-319)\\
\footnotesize 9&\footnotesize117(321-322) &  \footnotesize  24(113-215)\\
\footnotesize 10&\footnotesize5(102-106) &  \footnotesize  107(316-317)\\
\footnotesize 11&\footnotesize83(302-304) &  \footnotesize  41(123-217)\\
\footnotesize 12&\footnotesize79(221-222) &  \footnotesize  69(216-217)\\
\hline
\hline
\end{tabular}\label{tb:96cross}
\end{table}

\section{Conclusions}
This paper proposes a SF-network method to identify the critical components in cascading failures.
The SF-network is formed by cascading failure chains that contains the causal relationships between component failures, and losses that measure the blackout severities.
It can be used to identify the critical failures and critical components in the propagation phase of cascading failures.

The validations verify that upgrading the components with higher criticality indexes can decrease more blackout risks.
Thus, the proposed method can contribute to power system reinforcements and expansion plans.

The criticality indexes of the components are obtained by failure risks, so relationships between component failures and blackout losses can be further analyzed in the future.
Besides, since the number of failure chains needed increases as the system scale gets larger, the methods to accelerate the simulation can be employed to improve the analytical efficiency of this method, which is promising and currently being pursued.

\section*{Acknowledgement}
This work is partly supported by State Grid Corporation of China (Project: The research and development of multi-sandpile theory based blackout early warning technologies and systems in interconnected power grids).

\bibliographystyle{model3-num-names}
\bibliography{Bibliography}







\end{document}